\documentclass[%
 preprint, 3p, twocolumn,
]{elsarticle}

\usepackage{graphicx}% Include figure files
\usepackage{dcolumn}% Align table columns on decimal point
\usepackage{bm}% bold math
\usepackage{float} 
\usepackage{ulem} 
\usepackage{xcolor}
\usepackage{amsmath, amsfonts}
\usepackage{hyperref}% add hypertext capabilities
\usepackage[mathlines]{lineno}% Enable numbering of text and display math
\makeatletter
\def\ps@pprintTitle{%
   \let\@oddhead\@empty
   \let\@evenhead\@empty
   \let\@oddfoot\@empty
   \let\@evenfoot\@oddfoot
}
\makeatother

\begin{document}

\title{Bioluminescence modeling for deep sea experiments}
\author[1]{Meighen-Berger S.\fnref{fn1}}
\author[3]{Ruohan L.\fnref{fn3}}
\author[2]{Wimmer G. A.\fnref{fn2}}

\fntext[fn1]{stephan.meighen-berger@tum.de}
\fntext[fn2]{Current affiliation: Los Alamos National Laboratory, NM, USA; gwimmer@lanl.gov}
\fntext[fn3]{li.ruohan@physik.uni-muenchen.de}

\address[1]{Technische Universit\"at M\"unchen, James-Franck-Stra{\ss}e, 85748, Garching, Germany}
\address[2]{Imperial College London, London SW7 2AZ, UK}
\address[3]{Ludwig-Maximilians-Universit\"at M\"unchen, Schellingstra{\ss}e 4, 80799, M\"unchen, Germany}

\date{\today}% It is always \today, today,

\begin{abstract}
We develop a modeling framework for bioluminescence light found in the deep sea near neutrino telescopes by combining a hydrodynamic model with a stochastic one. The bioluminescence is caused by organisms when exposed to a non-constant water flow, such as past the neutrino telescopes. We model the flow using the incompressible Navier-Stokes equations for Reynolds numbers between 4000 and 23000. The discretization relies on a finite element method which includes upwind-stabilization for the velocity field. On top of the flow model, we simulate a population of random microscopic organisms. Their movement and emission are stochastic processes which we model using Monte Carlo methods. We observe unique time-series for the photon counts depending on the flow velocity and detector specifications. This opens up the possibility of categorizing organisms using neutrino detectors. We show that the average light-yield and pulse shapes require precise flow modeling, while the emission timing is chaotic. From this we construct a fast modeling scheme, requiring only a subset of computationally expensive flow and population modeling.
\end{abstract}

%\keywords{Supersymmetry, Cherenkov Telescopes, IceCube, Stau, Neutrino, Navier-Stokes}
\maketitle

%\tableofcontents
\section{Introduction}
Currently multiple new oceanic neutrino telescopes are under construction, such as P-ONE \cite{Agostini:2020aar} and KM3NET \cite{Adrian-Martinez:2016fdl}. Unlike IceCube-Gen2 \cite{vanSanten:2017chb} and its predecessor, Cherenkov telescopes in the ocean have to contend with an additional source of light, bioluminescence. Bioluminescence is optical light produced chemically by organisms in form of either flashes or a steady glow. The spectral distributions of these pulses peak in the range 450 nm to 490 nm \cite{latz_spectral_1988, doi:10.1098/rspb.1983.0095}, which is near to the peak of the expected Cherenkov spectrum at 420 nm \cite{CLAUS1987540, kulcsar_study_1982}. This makes disentangling the two sources difficult. Studies from ANTARES \cite{Aslanides:1999vq} and KM3Net show that neutrino telescopes are capable of measuring bioluminescence \cite{aguzzi_inertial_2017, article_badr}. While estimations of bioluminescence in neutrino telescopes have been done previously \cite{priede_potential_2008, CRAIG2009224}, these were done mainly focused on emission due to encounters. Their rate was found to be $\approx 1$ h$^{-1}$. Here we model emissions due to encounters and shear stress. The latter of which has a potential rate of $\approx 1-2$ magnitudes larger, see Figure \ref{fig:counts_vs_density}. This makes precise water flow modeling around the detector a requirement. Previously such studies were done for specific cases of bioluminescence such as dolphin-stimulated \cite{Rohr1447} and nozzle throats \cite{Latz1941}. Here we construct a realistic bioluminescence model based on a water current simulation and a Monte Carlo model for the organisms. This framework can be applied to any desired geometry. The framework for the model is called \textit{Fourth Day}\footnote{\url{https://github.com/MeighenBergerS/fourth_day}} and is publicly available.

\section{Bioluminescence}
Two external sources causing organisms to flash are given by contact and shear forces. The first of which is a response to the organism colliding with another object. The second source, shear stress, is a defensive response to an applied shear force, such as water turbulence \cite{CRAIG2009224, article_rohr_1998, https://doi.org/10.4319/lo.1999.44.6.1423, nauen_biomechanics_1998, ROHR20022009, https://doi.org/10.1029/2003JC001871}. This response has been observed in background currents as low as 5 mm/s \cite{doi:10.2307/1542610}, far lower than the cm/s velocities seen in deep ocean experiments. The emission probability due to shear stress follows a binomial distribution, with a successful emission probability denoted by $P_\mathrm{shear}$. The emission probability is linearly related to the water current's gradient \cite{10.2307/25066606, Rohr1447, bookSchlichting, laufer_structure_1954}
    \begin{equation}\label{eq:shearprob}
        P_\mathrm{shear} \propto \alpha \nabla\mathbf{u}.
    \end{equation}
Here $\mathbf{u}$ describes the flow velocity and $\alpha$ is a proportionality factor. For simplicity we assume $\alpha$ is a species independent constant. These shear emissions make precise water current modeling a necessity, which we describe in Section (\ref{sec:section_modelling}). Another source of flashes are spontaneous emissions. These are rare, due to the high energy cost for the organisms \cite{widder_bioluminescence_1989} and for this reason we neglect it. Once an organism flashes, be it due to an encounter or shear force, it will produce approximately $10^9 - 10^{13}$ photons \cite{priede_potential_2008}. These photons are then attenuated while traveling to the detector, as described in Section (\ref{sec:light_prop}).

\section{Population Modeling}\label{sec:pop}
    The organisms are modeled using a Monte Carlo scheme. Each individual is assigned a position and species. According to the species, the following set of properties is defined as probability distribution functions (pdf): Spectral emission \cite{latz_spectral_1988, doi:10.1098/rspb.1983.0095}, photon count, duration of emission, depth and movement \footnote{For a detailed description of the pdfs please check the data files in: \url{https://github.com/MeighenBergerS/fourth_day}}. These distributions are then sampled when required. Both the density of organisms and the probability coefficient $\alpha$ in equation (\ref{eq:shearprob}) are treated as unknowns. For sites such as ANTARES the density is $\approx 0.02\;\mathrm{m}^{-3}$ \cite{priede_potential_2008}, which can be used as a rough estimate. At each time-step the velocity and position of the organisms are updated according to the external flow velocity field together with their own sampled velocity distributions. Organisms that encounter an object -- such as a detector -- are assumed to emit light. If a given organism is not flashing due to an encounter, the binomial distribution with the emission probability from equation (\ref{eq:shearprob}) is sampled. On a successful flash, the duration of the flash is again sampled from another species dependent pdf. Additionally, we assign a maximum emission energy to each organism, which is reduced when an emission occurs and slowly regenerates over time.

\section{Current Modeling} \label{sec:section_modelling}
\begin{figure}[ht]
\begin{center}
\includegraphics[width=0.45\textwidth]{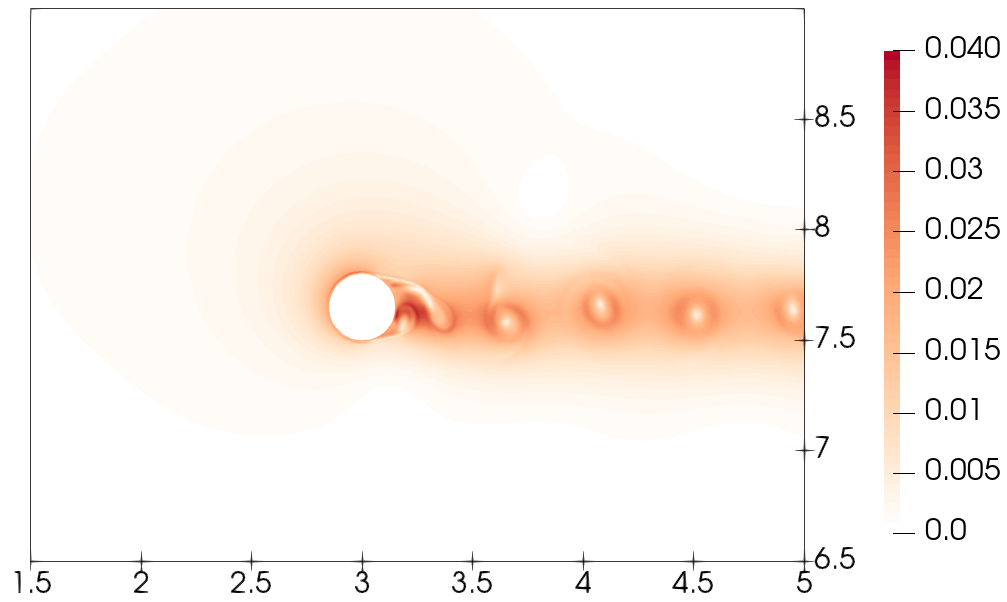}
\caption{Magnitude of relative velocity flow field $\mathbf{u}(x, t) - \mathbf{u}(x, 0)$ after $t = 300$ seconds for $u_{in} = 2$ cm/s, zoomed to circle; scale in cm/s. Axes labels are given in metres. The boundary layer is separated from the circle, generating vortices which form a vortex street.} \label{flow_past_cylinder}
\end{center}
\end{figure}

A typical shape of submerged neutrino detectors is given by a spherical or vertically arranged cylindrical body with a round top and bottom. In order to model the flow past such detectors numerically, we consider for simplicity a two-dimensional horizontal cross-section of the problem. This way, we aim to obtain a realistic approximation to the flow at least past the detector's cylindrical middle part, noting that for the spherical top and bottom parts a full three-dimensional setup would be required. The model described in this section can in principle be extended readily to the three-dimensional case, albeit at the expected substantial increase in computational cost.\\ \\
The flow is governed by the incompressible Navier-Stokes equations, which are given by a momentum equation and incompressibility condition of the form
\begin{align}
    &\partial_t \mathbf{u} + (\mathbf{u} \cdot \nabla) \mathbf{u} + \nabla p - \nu \Delta \mathbf{u} = 0, \label{NS_eqn_momentum}\\
    &\nabla \cdot \mathbf{u} = 0, \label{NS_eqn_incompr}
\end{align}
for flow velocity $\mathbf{u} = (u(\mathbf{x}), v(\mathbf{x}))$ and pressure $p(\mathbf{x})$, where $\mathbf{x} = (x, y)$ denotes the space coordinate. $\partial_t$ denotes the partial derivative with respect to time $t$, $\nabla = (\partial_x, \partial_y)$ the gradient, and further $\Delta = \partial_{xx} + \partial_{yy}$ denotes the diffusion operator. Finally, $\nu = \mu/\rho$ is given by the kinematic viscosity, for viscosity $\mu$ and density $\rho$. In the following, we will set $\mu = 0.001306$ Ns/m$^2$ and $\rho = 999.7$ kg/m$^3$, which corresponds to water at 10 degrees Celsius. Note that in terms of the flow's qualitative behaviour past the cylinder, we expect similar results for the typical temperature range of sea water\footnote{In terms of the Reynolds number for a $10$ cm/s flow as described below, we have $Re \approx 16,000$ for $0^{\circ}$C, and $Re \approx 30,000$ for $20^{\circ}$C, which leads to the same turbulent regime as for the Reynolds number  $Re \approx 23,000$ corresponding to $10^{\circ}$C.}.\\ \\
We consider a rectangular domain $\Omega$ of length $L_x =30$ m and width $L_y=15$ m. The domain includes a circular gap corresponding to the detector, which has a radius of $15$ cm and is placed at $\mathbf{x}= (3, 7.65)$ m. The cylinder's slight off-set in the $y$-direction is complemented with a parabolic inflow boundary condition at the domain's left edge, which is given by
\begin{align}
    &\mathbf{u}|_{x=0} = \left(u_{in} + u_p\left(\frac{4y}{L_y^2}(L_y - y) - 1\right), 0\right). \label{inflow_bc}
\end{align}
In the following, we will set maximum inflow velocities (occurring at $y=L_y/2$) according to
\begin{equation}
    u_{in} \in \{2\text{ cm/s}, 5 \text{ cm/s}, 10 \text{ cm/s}\},
\end{equation}
and fix $u_p = 0.5$ cm/s. The parabolic inflow, together with the detector's offset, leads to a non-symmetric initial flow profile (with respect to the detector), which ensures the formation of a K\'arm\'an vortex street as discussed further below in this section. \\ \\
To complete the model's description, we additionally require the remaining boundary conditions as well as a set of initial conditions. The former are given by normal flow conditions of the form
\begin{align}
    &\mathbf{u}|_{y \in \{0, L_y\}} = (u_{in} - u_p, 0), \label{wall_bc}\\
    &\mathbf{u}|_{\text{at detector}} = (0, 0), \label{cylinder_bc}\\
    &p|_{x = L_x} = 0. \label{outflow_bc}
\end{align}
We will find these restrictions to be sufficient for our purposes of simulating the flow, and in particular no boundary condition is set e.g. for $\mathbf{u}$ at the outflow boundary. Finally, we consider initial conditions of the form
\begin{equation}
    \mathbf{u}(\mathbf{x}, 0) = \mathbf{u}|_{x=0}, \;\;\; p(\mathbf{x}, 0) = \frac{8\nu u_{in}}{L_y^2}\frac{L_x - x}{L_x},
\end{equation}
which satisfy the Navier Stokes equations in a steady state for the rectangular domain without the cylindrical gap corresponding to the detector, and given the specified boundary conditions. Note that the flow will adjust to the detector after the first time step in the numerical model.\\ \\
The problem of flow past a circle has been studied extensively in literature (e.g. \cite{engelman1990transient, lienhard1966synopsis, norberg2003fluctuating, roshko1953development, wille1960karman}), and its complexity depends on the inflow speed $u_{in}$, kinematic viscosity $\nu$, and circle diameter $d$. These three parameters can be combined to the non-dimensional Reynolds number
\begin{equation}
    Re = \frac{u_{in}d}{\nu},
\end{equation}
which can then be used to determine the type of flow \cite{lienhard1966synopsis}. In our case, the inflow velocities lead to
\begin{align}
    &u_{in} = 10 \text{ cm/s} \; \implies \;\; Re \approx 23,000,\\
    &u_{in} = 5 \text{ cm/s} \;\;\; \implies \;\; Re \approx 11,500,\\
    &u_{in} = 2 \text{ cm/s} \;\;\; \implies \;\; Re \approx 4,600,
\end{align}
and we expect to obtain a boundary layer -- i.e. a thin layer in which the velocity decreases rapidly to the prescribed zero value at the boundary -- around the circle, which is separated from the circle towards the circle's wake (see Figure \ref{flow_past_cylinder}). The layer's thickness is proportional to $\sqrt{\nu}$ \cite{elman2014finite}, and for the above Reynolds numbers, the layer is laminar (i.e. non-turbulent). Pairs of trailing vortices are created and shed periodically between the separated boundary layer filaments near the circle's wake. The vortex shedding in turn creates a so-called K\'arm\'an vortex street of pairs of vortices in the circle's wake \cite{wille1960karman}. The periodicity of this process is represented by the non-dimensional Strouhal number $St = f d/u_{in}$, where $f$ denotes the frequency at which vortex pairs are shed. $St$ can be formulated as a function of $Re$, and for $Re \approx 23,000$, we expect $St \approx 0.2$ \cite{lienhard1966synopsis}. \\ \\
Finally, we note that for Reynolds numbers between $300$ and $30,000$, the vortex street is turbulent, in the sense that the vortex pairs interact strongly with each other and degrade the periodic pattern along the flow in the circle's wake (see Figures \ref{flow_past_cylinder}, \ref{flow_at_300}). This holds true in particular for Reynolds numbers above $10,000$, and for the above choices of flow speed, we expect the $5$ cm/s and $10$ cm/s flows to exhibit a strongly degraded pattern. In contrast, for the $2$ cm/s flow, we expect the degradation to be less pronounced.
\subsection{Discretization}
\begin{figure}[ht]
\begin{center}
\includegraphics[width=0.45\textwidth]{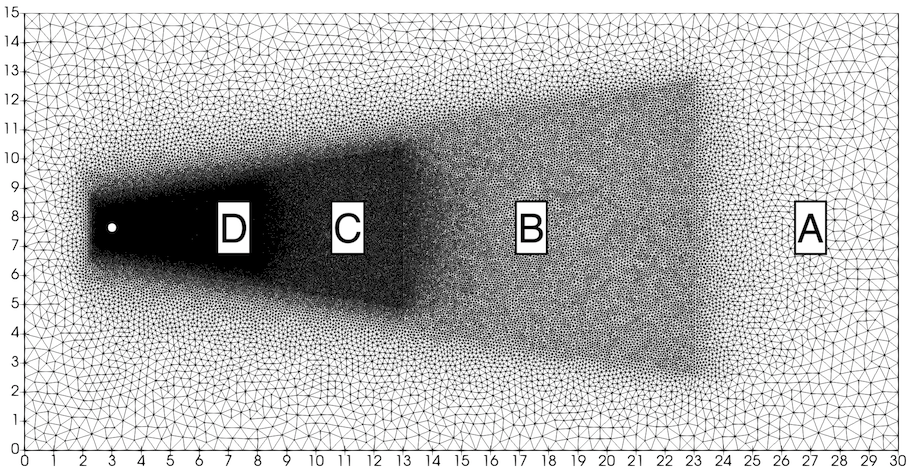}
\caption{Mesh used for flow simulation. Regions A, B, C, and D correspond to a resolution of $50$ cm, $10$ cm, $5$ cm, and $2.5$ cm, respectively. Near the circle, another refined region corresponds to a resolution of $1$ cm (see Figure \ref{mesh_E}). x and y axes are given in metres.} \label{mesh_A_to_D}
\end{center}
\end{figure}

\begin{figure}[ht]
\begin{center}
\includegraphics[width=0.45\textwidth]{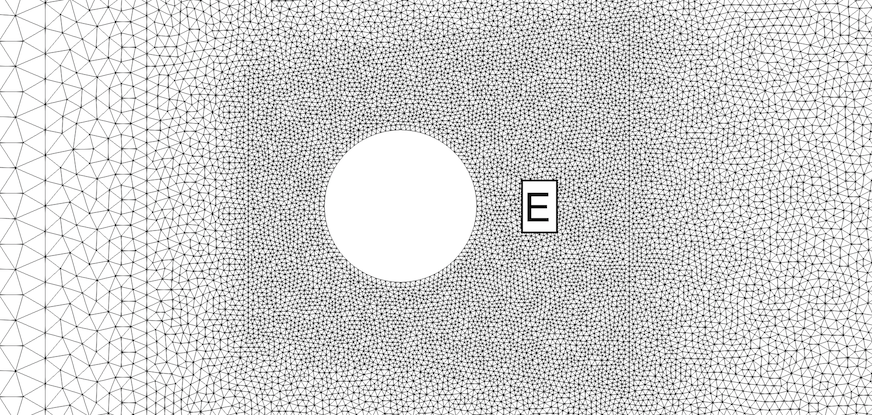}
\caption{Mesh used for flow simulation, zoomed to region near gap corresponding to detector. Region E corresponds to a resolution of $1$ cm.} \label{mesh_E}
\end{center}
\end{figure}

\begin{figure*}[ht]
\begin{center}
\includegraphics[width=1.0\textwidth]{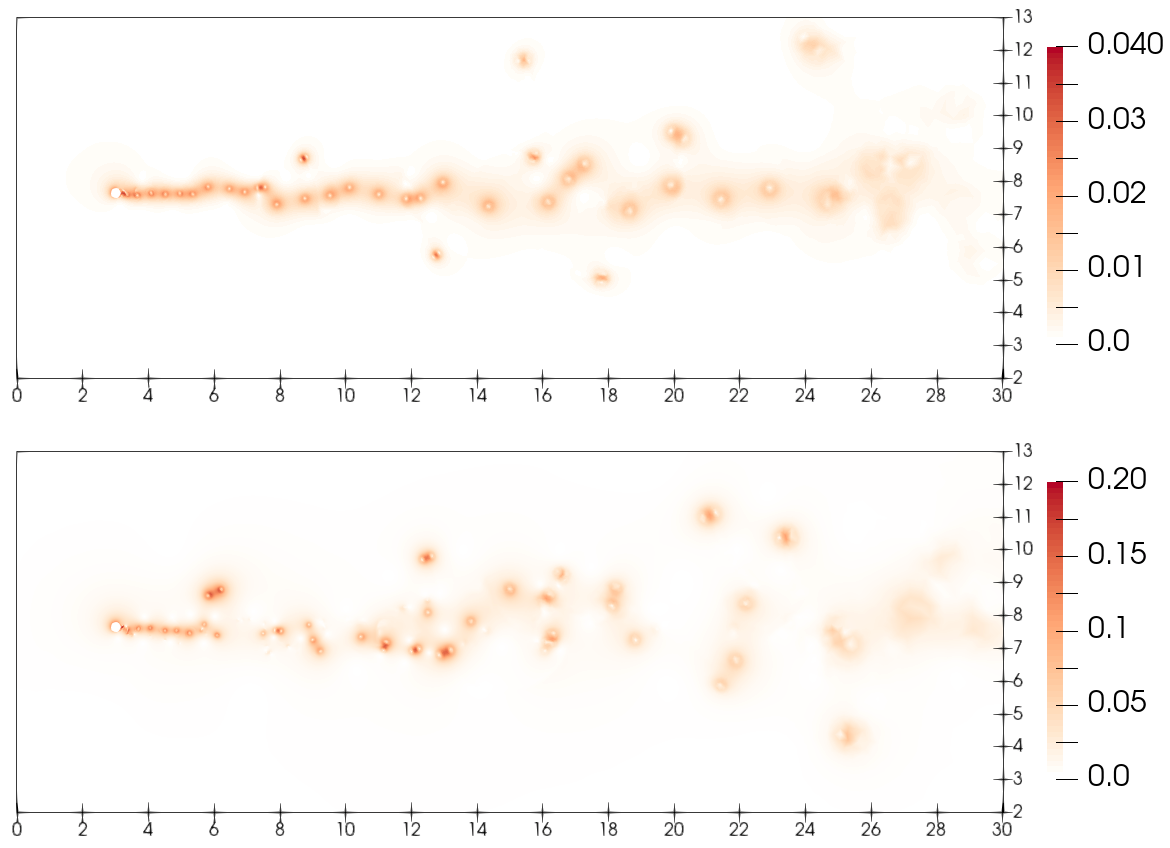}
\caption{Magnitude of relative velocity flow field $\mathbf{u}(x, t) - \mathbf{u}(x, 0)$ after $t = 300$ seconds for $u_{in} = 2$ cm/s (top) and $u_{in} = 10$ cm/s (bottom); scale in m/s. For the Reynolds numbers considered here, the vortex street is turbulent, leading to a degraded periodic pattern. At $x=23$ m, the vortices enter the coarse mesh region and are damped more strongly by the SUPG stabilization method. Axes labels are given in metres.} \label{flow_at_300}
\end{center}
\end{figure*}

We discretize the model in space using the finite element method. This way, we can readily use irregular meshes, which facilitate the domain's discretization near the circular gap corresponding to the detector, and which further allows for a varying resolution in the domain. In particular, for a higher efficiency in terms of computational cost, we employ a relatively fine resolution ($1$ cm) near the vortex shedding area, a slowly decreasing resolution along the circle's wake ($2.5$ cm, $5$ cm, and $10$ cm), and a coarse resolution ($50$ cm) away from the circle and wake (see Figures \ref{mesh_A_to_D}, \ref{mesh_E}).\\
We consider Taylor-Hood finite elements for the velocity and density field \cite{taylor1973numerical}, and further apply an Incremental Pressure Correction Scheme (IPCS, \cite{goda1979multistep}) for the time discretization. Note that the above choice of resolutions in the domain's various regions leads to under-resolved areas; in particular, this includes the boundary layer at the circle, as well as vortices further along the wake. To avoid the corresponding instabilities \cite{quarteroni2010numerical}, we also include a Streamline Upwind Petrov Galerkin (SUPG, \cite{brooks1982streamline}) method in the discretized velocity equation.\\ \\
The velocity solution space $\mathbb{V}_u$ is given by the space of vector-valued continuous piece-wise quadratic polynomials, and the pressure solution space $\mathbb{V}_p$ consists of scalar continuous piece-wise linear polynomials. Both of these spaces are further assumed to be restricted to functions that satisfy the boundary conditions \eqref{inflow_bc} and \eqref{wall_bc} - \eqref{outflow_bc}. The momentum equation \eqref{NS_eqn_momentum} is space-discretized according to
 \begin{align}
     \langle \mathbf{w}, \partial_t \mathbf{u} \rangle &+ \langle \mathbf{w}, (\mathbf{u} \cdot \nabla) \mathbf{u} \rangle \nonumber\\
     &+ P(\mathbf{u}, p; \mathbf{w}) \label{discr_u_eqn}\\
     &+ \langle \tau (\mathbf{u} \cdot \nabla) \mathbf{w}, \mathbf{u}_{res} \rangle= 0 & \forall \mathbf{w} \in \mathring{\mathbb{V}}_u, \nonumber
 \end{align}
 where $\langle.,. \rangle$ denotes the $L^2$-inner product, and $\mathring{\mathbb{V}}_u$ denotes the velocity space with functions vanishing at the boundary. The term in the second line of \eqref{discr_u_eqn} corresponds to the discrete Cauchy stress tensor
 \begin{equation}
     \sigma(\mathbf{u}, p) = 2 \nu \epsilon(\mathbf{u}) - pI,
 \end{equation}
 for symmetric gradient $\epsilon(\mathbf{v}) = (\nabla \mathbf{v} + \nabla \mathbf{v}^T)/2$ and $2\times2$ identity matrix $I$, and is given by
 \begin{equation}
     P(\mathbf{u}, p; \mathbf{w}) = \langle \sigma(\mathbf{u}, p), \epsilon(\mathbf{w}) \rangle.
 \end{equation}
 Finally, the term in the third line of \eqref{discr_u_eqn} corresponds to the SUPG stabilization, and the discretized momentum equation's residual appearing in it is given by
 \begin{equation}
     \mathbf{u}_{res} = \partial_t \mathbf{u} + (\mathbf{u} \cdot \nabla) \mathbf{u} - \nabla \cdot \sigma(\mathbf{u}, p).
 \end{equation}
 Further, the stabilisation parameter is set to
 \begin{equation}
     \tau = \frac{1}{2} \! \left(\!\left(\frac{2\kappa(x)}{\Delta t}\right)^{\!2} \!\!\!+\! \left(\frac{2|\mathbf{u}|}{\Delta x}\right)^{\!2} \!\!\!+\! 9\left(\frac{4\nu}{\Delta x^2}\right)^{\!2}\right)^{-\frac{1}{2}}\!\!\!\!, \label{def_tau}
 \end{equation}
 where $\kappa(x) = 1$ for $x<3.75$ m, and $\kappa(x) = 1/6$ otherwise. $\Delta t$ and $\Delta x$ denote the time step and (local) mesh size, respectively. This choice of $\tau$ is similar to one often considered in the literature (e.g. \cite{tezduyar1991stabilized}), up to the additional parameter $\kappa$. In the model used in this paper two sources of instabilities appeared frequently: first, towards the later stages of the wake, where the resolution is relatively coarser, vortices may become unstable. Second, the separated boundary layer filaments may become unstable. We found that different values of $\kappa(x)$ were suited best for these two types of instability, thus motivating the split used in $\tau$ as given above.\\ \\
 Before moving on to the time discretization, we include a small remark on the significance of $\tau$. For this purpose, we ignore the viscosity contribution in \eqref{def_tau}, set $\kappa(x) = 1$ and approximate the velocity $\mathbf{u}$ via $\Delta x$ and $\Delta t$ according to a CFL type criterion. The resulting SUPG stabilization parameter then reduces to
 \begin{equation}
     \tau \approx \frac{\Delta t}{4\sqrt{2}}.
 \end{equation}
With this in mind, the SUPG method can also be seen as a sub-grid scale model, and $\tau$ -- which is paired with the velocity residual $\mathbf{u}_{res}$ -- corresponds to a time scale in which sub-grid contributions take an effect on the resolved scales.
 \\ \\
 Next to the momentum equation, the incompressibility condition \eqref{NS_eqn_incompr} is used to derive an equation for the pressure field. The latter equation then appears in the time-discretized scheme, where the momentum equation is first solved for to obtain a predicted velocity $\mathbf{u}^p$. Given the nonlinear advection term $(\mathbf{u} \cdot \nabla) \mathbf{u}$, this is done in an iterative manner according to
 \begingroup
 \addtolength{\jot}{3mm}
 \begin{align}
     &\text{do } k = 1, ..., m \colon \nonumber\\
     &\hspace{3mm}\langle \mathbf{w}, (\mathbf{u}^{p,i+1} - \mathbf{u}^n)/\Delta t \rangle + \langle \mathbf{w}, (\bar{\mathbf{u}}^* \cdot \nabla) \bar{\mathbf{u}}^{p,i+1} \rangle \nonumber\\
     &\hspace{3mm}\;\;+ P(\bar{\mathbf{u}}^{p,i+1}, p^n; \mathbf{w}) \label{predictor_solve}\\
     &\hspace{3mm}\;\; + \langle \tau(\mathbf{u}^n) (\bar{\mathbf{u}}^* \cdot \nabla) \mathbf{w}, \bar{u}_{res}) \rangle= 0 \hspace{1cm} \forall \mathbf{w} \in \mathring{\mathbb{V}}_u, \nonumber
 \end{align}
 \endgroup
 where $(\mathbf{u}^n, p^n)$ are the known fields of the current time step (starting from the initial conditions $(\mathbf{u}^0, p^0)$). Further, the mid-point averages $\bar{\mathbf{u}}^{p,i+1}$ and $\bar{\mathbf{u}}^*$ are given by
 \begin{align}
     &\bar{\mathbf{u}}^{p,i+1} = (\mathbf{u}^n + \mathbf{u}^{p, i+1})/2,\\
     &\bar{\mathbf{u}}^*= (\mathbf{u}^n + \mathbf{u}^{p, i})/2,
 \end{align}
 for unknown $\mathbf{u}^{p,i+1}$ to be solved for, and the current iteration's known guess $\mathbf{u}^{p, i}$. Note that in the first iteration, the latter guess is set to $\mathbf{u}^n$, while in each following one it is set to the velocity field computed in the previous iteration. Finally, we note that the residual in the SUPG term is time-discretized according to
 \begin{equation}
 \begin{split}
     \bar{u}_{res} = &(\mathbf{u}^{p, i+1} - \mathbf{u}^n)/\Delta t + (\bar{\mathbf{u}}^* \cdot \nabla) \bar{\mathbf{u}}^{p, i+1}\\
     &- \nabla \cdot \sigma(\bar{\mathbf{u}}^{p, i+1}, p^n).
 \end{split}
 \end{equation}\\
 Once the final predicted velocity $\mathbf{u}^p$ has been computed in the loop's last iteration, the pressure corresponding to the next time level $n+1$ is obtained by solving an update equation of the form
 \begin{align}
     \Delta t \langle \nabla (p^{n+1} - p^n), \nabla q \rangle =- \langle \nabla \cdot \mathbf{u}^p, q \rangle,
 \end{align}
 for any test function $q \in \mathring{\mathbb{V}}_p$, where $\mathring{\mathbb{V}}_p$ is defined analogously to $\mathring{\mathbb{V}}_u$. Finally, given $(\mathbf{u}^p, p^{n+1})$, we compute a velocity field $\mathbf{u}^{n+1}$ according to
 \begin{align}
     \langle \mathbf{u}^{n+1} - \mathbf{u}^p, \mathbf{w} \rangle = - \Delta t \langle\nabla(p^{n+1} - p^n), \mathbf{w} \rangle,
 \end{align}
 for any test function $\mathbf{w} \in \mathring{\mathbb{V}}_u$.\\ \\
The simulations' time parameters are given by
\begin{align}
    &u_{in} = 2\text{ cm/s} \colon \;\; \Delta t = 0.05, \; t_{max} = 3000 s,\\
    &u_{in} = 5\text{ cm/s} \colon \;\; \Delta t = 0.02, \; t_{max} = 1200 s,\\
    &u_{in} = 10\text{ cm/s} \colon \; \Delta t = 0.01, \; t_{max} = 600 s,
\end{align}
noting that for each configuration, the flow covers a total distance of $60$ m $= 2L_x$. The mesh is implemented using Gmsh \cite{geuzaine2009gmsh}, and the finite element discretization is based on the automated finite element toolkit Firedrake\footnote{for further details, see \cite{homolya2018tsfc, luporini2017algorithm} or \url{http://firedrakeproject.org}} \cite{rathgeber2016firedrake}. The latter uses the solver library PETSc (see e.g. \cite{balay2019petsc, balay1997efficient}) to compute the resulting systems of equations. A snapshot of the simulation at 300 seconds is given in Figure \ref{flow_at_300}, depicting the turbulent vortex street's degraded periodic pattern. We find that as expected, the $10$ cm/s flow generates a more degraded periodic pattern than the $2$ cm/s flow. However, the Strouhal number is higher than anticipated, noting that for such turbulent flows many aspects of the numerical scheme impact the latter number. This includes subtle factors such as the distance of the side walls to the cylinder \cite{behr1995incompressible}, the inflow's offset relative to the cylinder, and the precise mesh setup near the cylinder.\\ \\
%%% In future work, add low Re flow simulation to confirm correct Strouhal number for non-turbulent flow
In our simulation runs, we found $m=2$ nonlinear iterations to be sufficient. Further, in each iteration, the predictor problem \eqref{predictor_solve} -- which amounts to the majority of the scheme's computational cost - is solved for using GMRES, with BoomerAMG \cite{falgout_hypre_2002} used as preconditioner.\\ \\
Finally, given the simulations, compressed and filtered versions of the velocity field as well as the magnitude of its gradient, i.e. $|\nabla \mathbf{u}|(\mathbf{x}, t)$, are saved every $100$ time steps. This corresponds to $10$ cm distance covered by the flow for each of the three configurations. The filtering -- which can be seen as a post-processing step -- removes small scale oscillations in coarser mesh regions, which have no detrimental effect to the flow dynamics, but may spuriously trigger flashes for small choices of proportionality factor $\alpha$. The compression is done for the purpose of reducing the amount of data to be stored, and is achieved by interpolating the filtered velocity output field onto a first polynomial order vector Lagrangian finite element space. We note that Figures \ref{flow_past_cylinder} and \ref{flow_at_300}  depict the velocity's compressed and filtered version. The filtering procedure is detailed further in \ref{sec:app:apps_psi_filter}.

We provide three data sets, which we use later in the analysis for 2 cm/s, 5 cm/s and 10 cm/s flows \cite{wimmer_data_2021}.

\section{Light Propagation}\label{sec:light_prop}
    Once the population simulation has finished, the light from each organism's flashes are propagated to the detector.
    We attenuate the photon count using an attenuation factor, $A$, given by
    \begin{equation}
        A = \frac{\exp\left(-r\times l\right)}{4\pi r^2}.
    \end{equation}
    Here $r$ references the distance between the organisms, at the time of its emission, to the detector. For the wavelength dependent factor, $l$, we use values from \cite{Bradner1992AttenuationOL}. For a distance of 10 meters the total photon count drops by approximately a factor of 1000. Given this exponential attenuation, flashes beyond 30 m distance are usually irrelevant for a detector.

\section{Results and Discussion}
\begin{figure}[htb]
        \begin{center}
        \includegraphics[scale=1]{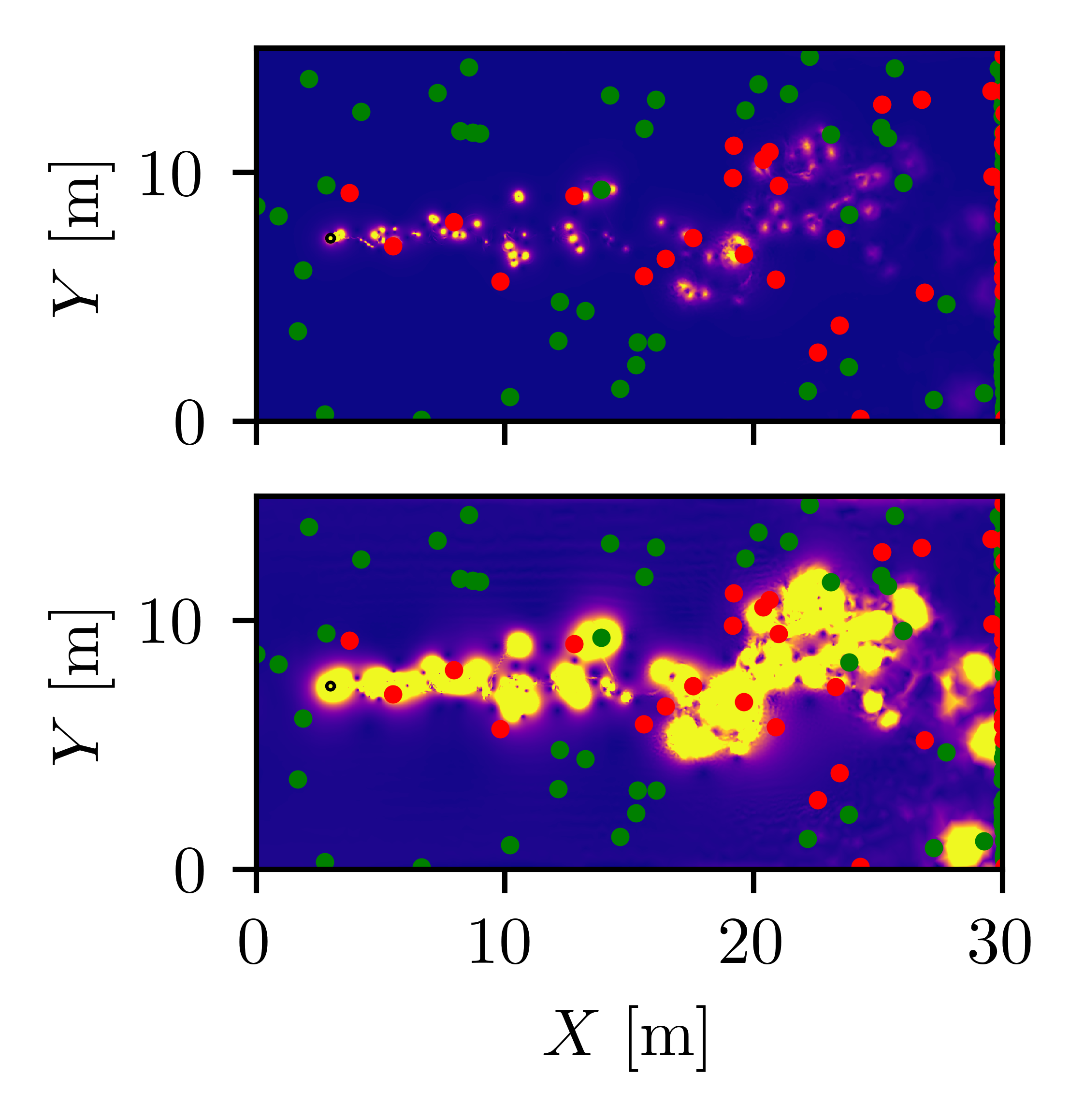}
        \caption{Distribution of organisms emitting light (red) and those that are not (green). The background shows the gradient of the velocity field. Clearly visible is the increase in emitting organisms in the detector's (black ring) wake. The top plot uses a larger range of gradients to visualize the vortices. The bottom one limits the range, emphasizing the regions where flashes due to shear stress are possible.}\label{fig:mc_pop}
        \end{center}
    \end{figure}

\begin{figure}[htb]
        \begin{center}
        \includegraphics[scale=1]{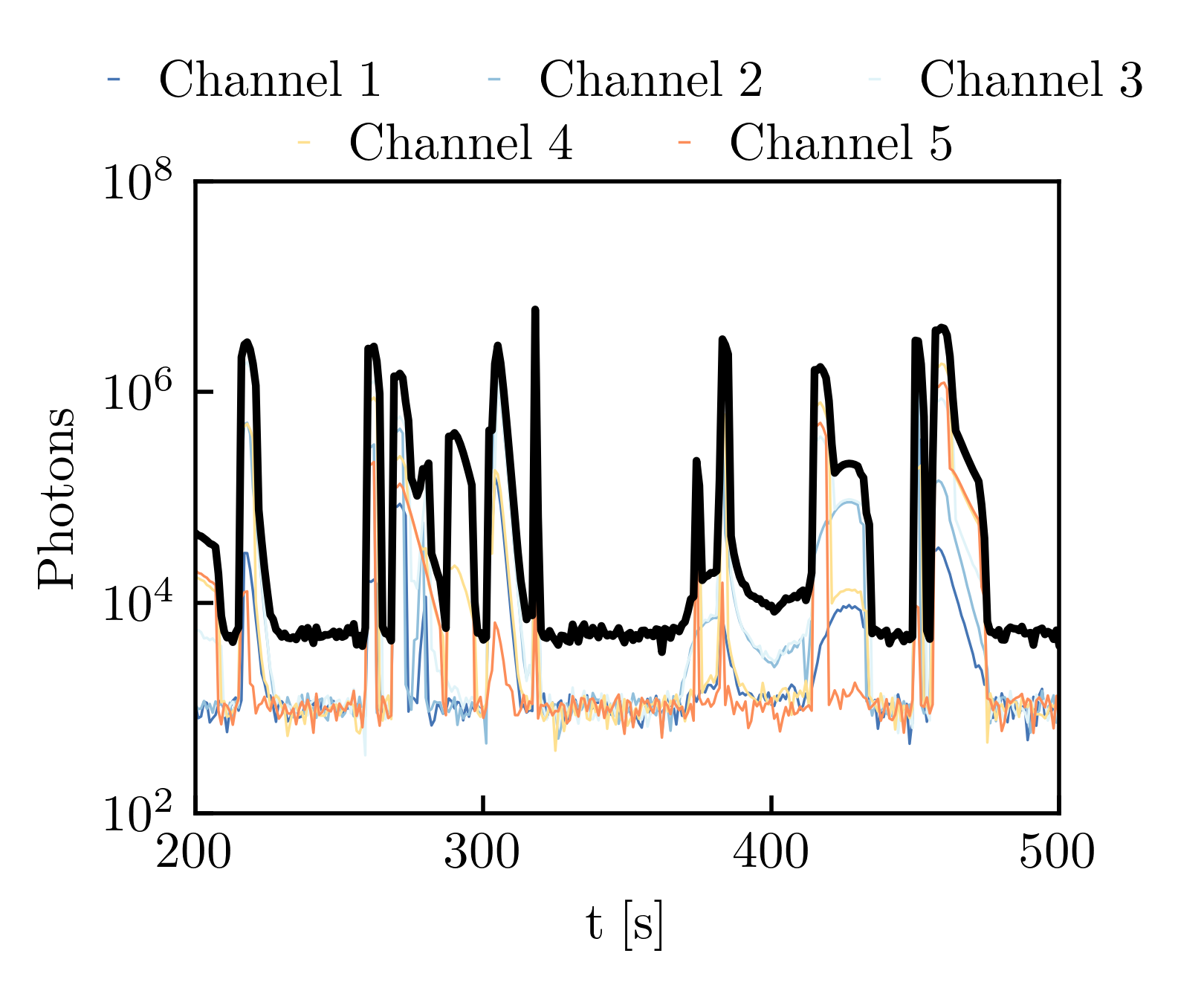}
        \caption{Example output for a spectrometer simulation with a 10 cm/s water current. Here we model five different PMTs measuring different wavelengths. In sequential order the wavelength ranges are [500 nm, 515 nm], [485 nm, 500 nm], [475 nm, 485 nm], [465 nm, 475 nm] and [455 nm, 465 nm]. The black line indicates the total photon count. Due to the finite opening angles of the detectors, here assumed to be 25$^\circ$, and the current velocity, single emission flashes show varying maxima positions for different wavelengths.}\label{fig:mc_example}
        \end{center}
    \end{figure}

\begin{figure}[htb]
        \begin{center}
        \includegraphics[scale=1]{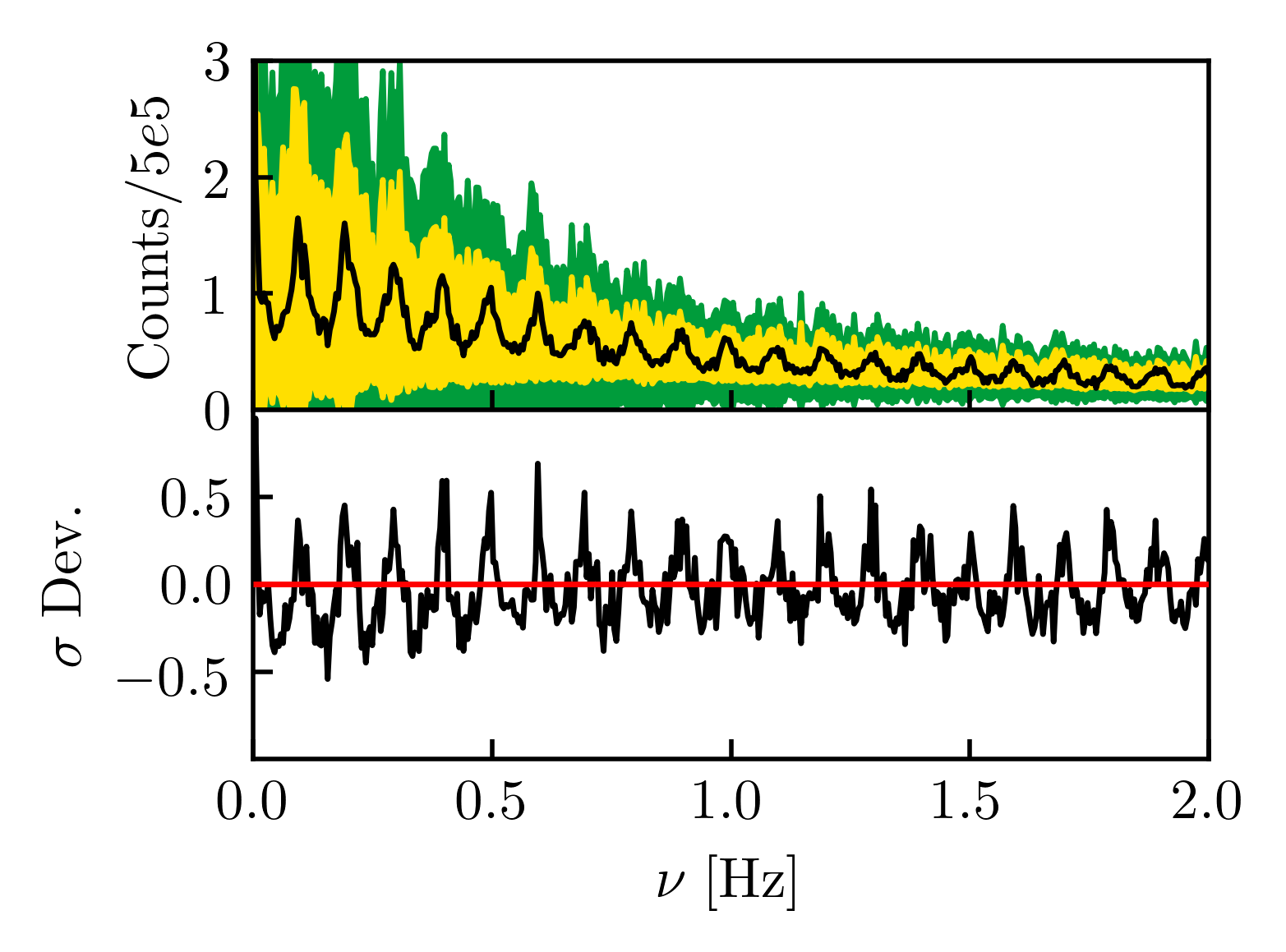}
        \caption{Fourier transform of the results from Figure \ref{fig:mc_example}. The top plot shows the spectrum of light reaching the detector. In black is the averaged spectrum from 100 simulation runs. In yellow and green we show the 1$\sigma$ and 2$\sigma$ confidence intervals of the background model. The bottom plot shows the deviation of the simulation from the null hypothesis in units of sigma.}\label{fig:mc_fft}
        \end{center}
    \end{figure}

\begin{figure}[htb]
        \begin{center}
        \includegraphics[scale=1]{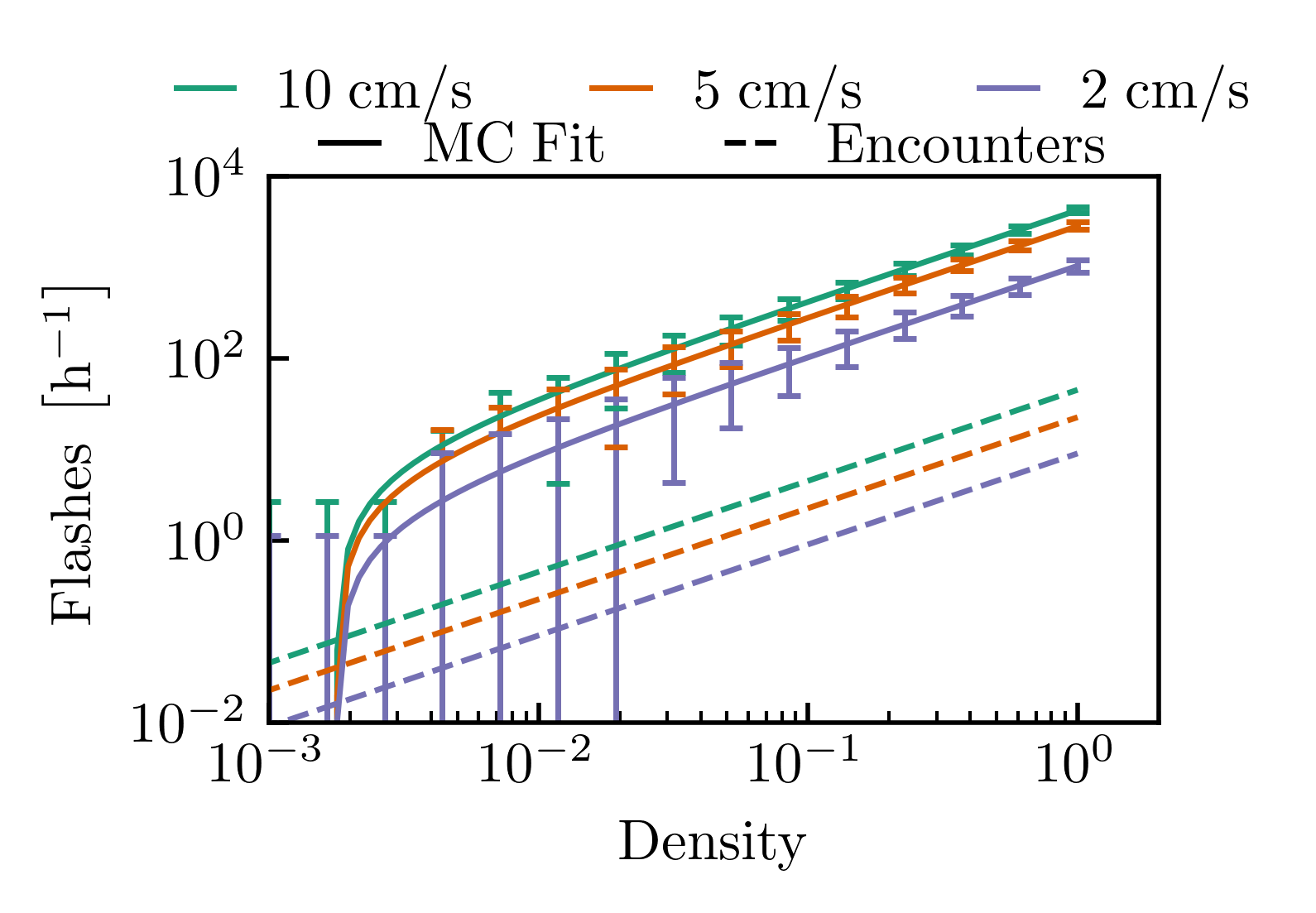}
        \caption{The number of flashes per hour as predicted by our simulation (dots) with corresponding linear fits (solid) and those predicted by encounters (dashed). Each simulation data point corresponds to the mean value from 16 h of simulated data. Here we set $\alpha=10$.}\label{fig:counts_vs_density}
        \end{center}
    \end{figure}

\begin{figure}[htb]
        \begin{center}
        \includegraphics[scale=1]{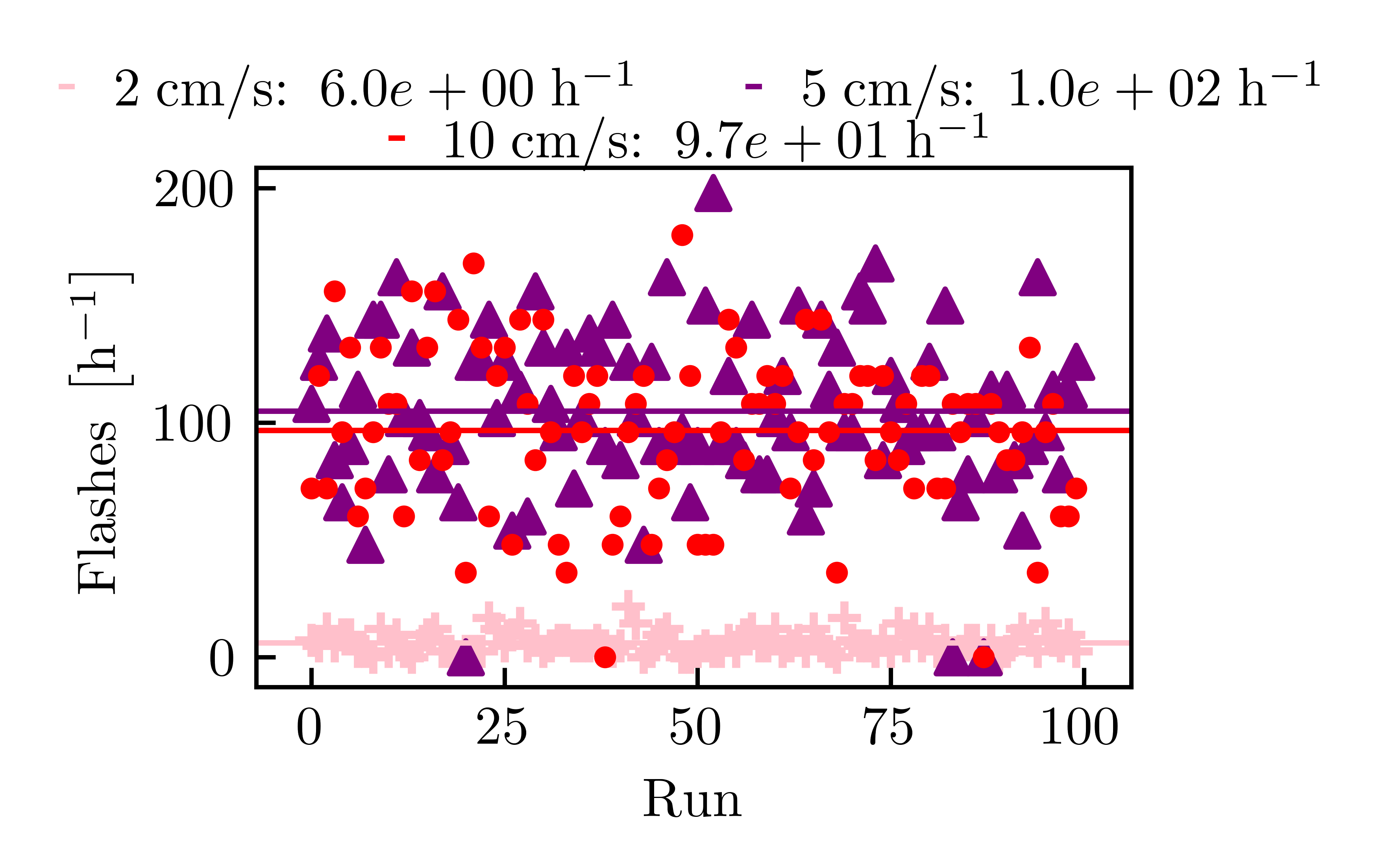}
        \caption{The velocity dependence of the number of flashes per hour. Here we show the results for each individual simulation run, corresponding to 10 minutes of simulated data. The lines represent the mean values over all runs. The crosses (pink), triangles (purple) and dots (red) represent the individual results for the 2 cm/s, 5 cm/s and 10 cm/s runs respectively.}\label{fig:vel_dependence}
        \end{center}
    \end{figure}

    Running a simulation for a perfect detector with a 360$^\circ$ field of view (FoV), we can confirm the expectations and estimates from \cite{priede_potential_2008, CRAIG2009224}, that most of the light is generated downstream of the module. As shown in Figure \ref{flow_at_300}, downstream the flow will be most turbulent and the wake wider, catching more organisms. Figure \ref{fig:mc_pop} emphasizes this point. In the top plot the range of the gradient was chosen to emphasize the vortices, while in the bottom plot to visualize the area where flashes due to shear stress are possible. In red we show currently flashing organisms, while in green the inactive ones.
    We can now introduce a more realistic detector, employing multiple photo multiplier tubes (PMTs), with limited FoVs, such as those of KM3Net, P-ONE and STRAW-b. Specifically we implement a model of the PMT Spectrometer employed by STRAW-b, which has a radius of 0.15 m\footnote{For a detailed description of the detector model please check the config file in: \url{https://github.com/MeighenBergerS/fourth_day}}. Assuming each PMT measures a specific wavelength range and possesses its own FoV results in Figure \ref{fig:mc_example}. There we modeled a 10cm/s water current, a population density of 0.02 $\mathrm{m}^2$ with $\alpha$ set to 10. The different channels denote different PMTs measuring different wavelength ranges. In black the total resulting time series is shown. Note the complex structure of some peaks. These are caused by an individual organism drifting from one detector's FoV to another, causing an apparent delay in the pulse between wavelengths. This is a unique type of signature, caused by the water current velocity and the detector's geometry. This means an individual organism can cause different measured time series, purely based on kinematics, disregarding any additional differences between pulses.
    Applying a Fourier transform to the simulation results leads to Figure \ref{fig:mc_fft}. There we constructed a background model by scrambling the pulses in time. The yellow and green bands show the one and two sigma deviations respectively for the background model. The unscrambled simulation (black) falls within the expectation of the scrambled one. This means the timing of the pulses is random. Due to the chaotic nature of the emission, we can construct realistic simulations by drawing from all possible pulse shapes and randomly distributing them according to the expected average, given in Figure \ref{fig:counts_vs_density}. This reduces the calculation time drastically, since the expensive population modeling can be skipped.
    As discussed in \cite{priede_potential_2008} the average number of expected encounter flashes can be modeled using the expected number organisms impinging on the detector per second. This rate is given by \cite{priede_potential_2008}
    \begin{equation}
        \mathrm{Impacts}^{-1} = \pi\left(r_\mathrm{detector} + r_\mathrm{animal}\right)^2\times u\times\rho.
    \end{equation}
    Here $r_\mathrm{detector} = 0.15\;\mathrm{m}$ and $r_\mathrm{animal}= 0.005\;\mathrm{m}$ are the radius of the detector and the mean radius of the animals respectively. The numerical values are the ones we used to compare to our simulation results. Plugging in the different water current velocities $u$ and different densities, given in organisms/m$^3$, results in the dashed lines in Figure \ref{fig:counts_vs_density}. There we also show the results of the simulation as dots and the corresponding linear fits (solid line). Note the large difference between the expected number of flashes from encounters and those predicted by our model, which includes shear emissions. From this we can deduce, that shear emission will play the dominant role when measuring time series. The fit values for this plot are given in Table \ref{tab:Fit_Vals}. Note the large differences between the 2 cm/s line and the others. This is due to the 2 cm/s scenario corresponding to a weakly turbulent regime with a less degraded periodic pattern when compared to the higher Reynolds number 5 cm/s and 10 cm/s setups.
    Rerunning the previous density simulation for $\rho = 0.1$ organisms / $\mathrm{m}^3$ and setting $\alpha=1$ results in Figure \ref{fig:vel_dependence}, where we show the results from each individual run. While the 2 cm/s run shows a low amount of variance, the other two velocities show a large amount of variance. Compared to Figure \ref{fig:counts_vs_density}, where we set $\alpha=10$, the mean number of flashes decreased by approximately a factor of 4 for the 5 cm/s and 10 cm/s cases, wile in the 2 cm/s case by 17. An additional observation is that the mean number of flashes between 5 cm/s and 10 cm/s only differ slightly in this regime. The higher mean for the 5 cm/s case may be explained by the running time of the flow model (with a distance of $2L_x$ covered by the flow). It corresponds to a relatively short snapshot, in which the 5 cm/s flow run may happen to overall exhibit a slightly more degraded pattern -- with more spread out vortex pairs within the wake -- than the 10 cm/s flow run. To fully cover the distribution corresponding to the degree to which the vortex pattern is degraded in time, considerably longer flow simulations would be required. Combined with the observed variance between simulation runs, differentiating between the 5 cm/s and 10 cm/s case proves difficult. Nonetheless, the increased variance is unlikely to explain the large difference between the 2 cm/s and the faster flow cases, which in turn fits well with the well-known difference in the extend of turbulence depending on the Reynolds number.\\
    
\begin{figure}[htb]
        \begin{center}
        \includegraphics[scale=1]{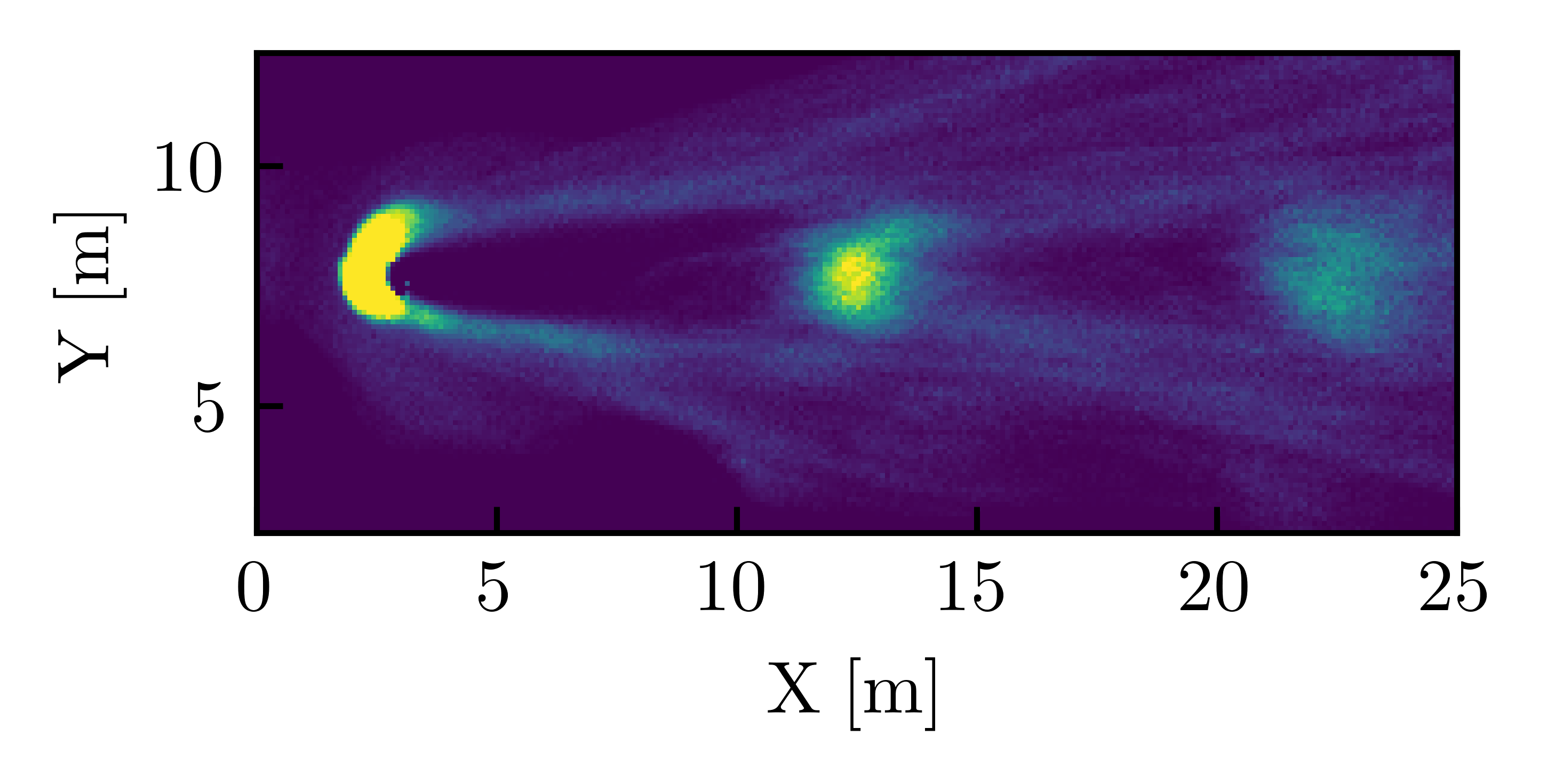}
        \caption{Depicted are the areas where the organisms start emitting light. Here we set the water current to 10 cm/s and $\alpha=5$. Note that the first blob corresponds to the detector location. This blob is followed by two distinct areas, where the organisms are far more likely to start flashing than others. This effect is barely visible in the semi-turbulant case. Additionally, the scattering of organisms is more prevalent here, than with the 2 cm/s flow. }\label{fig:3d_prob} %%% I think it is equally visible for the first three blobs of the 2cm/s flow plot; the real difference is the far bigger y-coordinate spread, no?
        \end{center}
    \end{figure}

    Finally, we generate a 167 h data set, setting $\alpha=5$. On this data set we analyze the point at which individual organisms start to flash. Figure \ref{fig:3d_prob} shows the resulting distribution. The color code runs from dark blue to yellow, with yellow the most likely point of emission. The first blob corresponds to the detector surface. Here the organisms first encounter the water current flowing around the cylinder, causing them to emit. These organisms then flow around the cylinder into its wake, while still emitting. It then takes some time for these organisms to emit again, or others to be dragged into the wake, which corresponds to the second blob. The third blob then follows a similar scheme to the second. This plot shows three distinct regions, where the likelihood of a pulse start lies far above the surrounding areas. These regions are observable in the turbulent case (5 cm/s and 10 cm/s) and less so in the semi-turbulent case (2 cm/s), see Figure \ref{fig:2cm_prob} in the appendix. Additionally, organism scattering is far less prominent in the semi-turbulent case, with all light being produced directly downstream of the detector.   Utilizing this prediction, studies on bioluminescence could be performed by studying these regions. On the other-hand, neutrino detectors, could filter their data for pulses coming from these regions, removing this background. %%% mention bigger y spread here; in the sense that this way a bigger area behind the cylinder is covered, therefore leading to flashes by organisms not directly in the wake, who would in the 2cm/s run not see a gradient.

\section{Conclusion}
    Predicting deep sea bioluminescence requires precise simulations of the water flow around detectors. Compared to encounter emission, flashes caused by shear stress play a far larger role at the low densities expected in the deep sea. We predict approximately 1-2 orders of magnitude higher rates for shear flashes than encounter emissions. For realistic detector modules, such as those employed by KM3Net, P-ONE and STRAW-b, we have identified unique bioluminescence signatures, defined by the flow velocity and the FoVs of the PMTs. These can in turn be used to identify the organisms drifting past these detectors, making deep sea detectors interesting for biologists as well. Finally we have demonstrated that while the emissions are chaotic in time, the starting location of the flashes are not for high flow velocities. These effects open unique opportunities for modeling and analyzing data gathered by neutrino detectors.

 \section{Acknowledgements}
 The material presented in this publication is based upon work supported by the Sonderforschungsbereich Neutrinos and Dark Matter in Astro- and Particle Physics (SFB1258). Golo Wimmer further acknowledges support from the EPSRC Mathematics of Planet Earth Centre for Doctoral Training at Imperial College London and the  University of Reading. Additionally, the research was supported by TU M\"unchen-Imperial College London collaboration fund. We thank Colin Cotter, Elisa Resconi and the P-ONE collaboration for their support.
 \appendix
 \section{Post-processing filter for current model output}\label{sec:app:apps_psi_filter}
Here, we briefly outline the filtering method used to remove small scale noise from the current model's velocity field output, which may otherwise contaminate the light flash analysis. The method is based on computing the field's stream function from the vorticity. For this purpose, given the velocity field $\mathbf{u}^n$ at a data output time $t = n\Delta t$, we compute the vorticity according to
 \begin{equation}
     \xi \in CG_1(\Omega)\colon \;\; \xi = P_{CG_1(\Omega)}\big(-\nabla \cdot {\mathbf{u}^n}^\perp\big),
 \end{equation}
 where $CG_1(\Omega)$ denotes the first order continuous Galerkin space defined on the domain, and $P_{CG_1(\Omega)}$ denotes the $L^2$-projection therein. Further, $^\perp$ denotes the perpendicular operator, i.e. ${\mathbf{u}^n}^\perp = (u_x, u_y)^\perp = (-u_y, u_x)$. The stream function $\psi \in CG_2(\Omega)$ -- for second order continuous Galerkin space with suitable boundary conditions determined such that $\nabla^\perp \psi$ is consistent with the boundary conditions of $\mathbf{u}^n$ -- is then solved for according to
 \begin{align}
     \langle \nabla \eta, \nabla \psi \rangle = \langle \eta, \xi \rangle && \forall \eta \in \mathring{CG}_2(\Omega).
 \end{align}
 Finally, the filtered velocity and the norm of its gradient are computed as projections of the form
 \begin{align}
     &\tilde{\mathbf{u}}^n = P_{\mathbb{V}_u}(\nabla^\perp \psi),\\
     &\widetilde{|\nabla\mathbf{u}^n|} = P_{\mathbb{V}_p}(|\nabla \tilde{\mathbf{u}}^n|).
 \end{align}
 
 \section{Linear Fits}
 Here we show the resulting fit values for Figure \ref{fig:counts_vs_density} for the different flow velocities. The values are given in Table \ref{tab:Fit_Vals} assuming a linear fit model $f(\rho)=k\rho+d$ with $\alpha = 10$.
 \begin{table}[htb]
 \begin{center}
\begin{tabular}{ |c|c|c| } 
 \hline
 Flow & $k$ & $d$ \\
 \hline
 10 cm/s & 4259 & -7.6 \\ 
 \hline
 5 cm/s & 2850 & -5.0 \\ 
 \hline
 2 cm/s & 1045 & -1.9 \\
 \hline
\end{tabular}\caption{The fit values for Figure \ref{fig:counts_vs_density}. Note that for the densities typically of interest, the differences between 10 cm/s and 5 cm/s is minimal and lie within the bounds of the simulation error, while it is large between 2 cm/s and the other two.}\label{tab:Fit_Vals}
\end{center}
\end{table}

\section{Emission Start 2 cm/s}
\begin{figure}[htb]
        \begin{center}
        \includegraphics[scale=1]{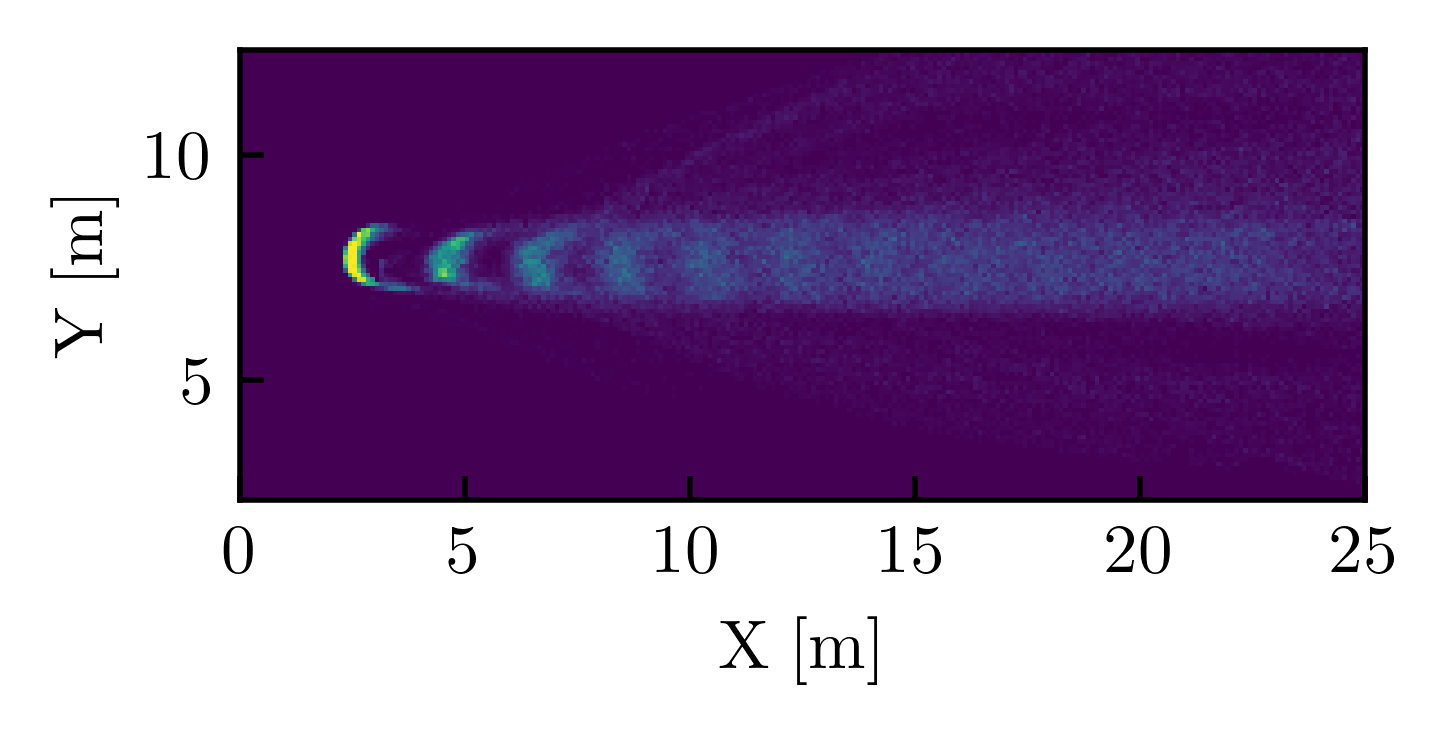}
        \caption{Depicted are the areas where the organisms start emitting light. Here we set the water current to 2 cm/s and $\alpha=5$. Note that the first blob corresponds to the detector location. Unlike the 10 cm/s the vortex emission angle is far less chaotic, leading to most light being produced behind the detector. Additionally distinct emission regions are far less prominent.}\label{fig:2cm_prob}
        \end{center}
    \end{figure}

As discussed in the main text, we tracked the starting position of each flash and constructed emission probabilities based on these. In the turbulent case, shown in Figure \ref{fig:3d_prob}, we observed three distinct regions, where the flashes are likely to start. In Figure \ref{fig:2cm_prob} we show the same plot for the 2 cm/s scenario. There, distinct regions, beside the one at the detector are barely visible. Additionally the vortex pair emission is far less chaotic, leading to most emission happening directly behind the detector.
 %
 %
 %\section{Generic Appendix Describing Boring Mathematics}\label{sec:app:boring_math}
 % Some boring mathematics
 \nocite{*}
\bibliographystyle{apsrev}
 \bibliography{bio}% Produces the bibliography via BibTeX.

\end{document}